\documentclass[10pt,conference]{IEEEtran}
  
\usepackage[square,numbers,sort&compress]{natbib}
\usepackage{booktabs}
\usepackage{seqsplit}
\usepackage{epsfig}
\usepackage[font=small]{caption}
\usepackage{subcaption}
\usepackage{multirow, makecell}
\usepackage{latexsym}       
\usepackage{amsmath}        
\usepackage{amssymb}
\usepackage{xspace}         
\usepackage[table]{xcolor}         
\usepackage{wasysym}        
\usepackage{tabularx}       
\usepackage{flushend}
\usepackage[utf8]{inputenc} 
\usepackage{rotating}
\usepackage{textcomp}
\usepackage[notextcomp]{stix}
\usepackage[shortlabels]{enumitem}

\usepackage{adjustbox}
\usepackage{array}

\renewcommand{\paragraph}[1]{\vspace{5pt}\noindent\textbf{#1.\xspace}}
\newcolumntype{R}{>{\raggedleft\arraybackslash}X}
\newcolumntype{L}{>{\raggedright\arraybackslash}X}

\newcommand{\todo}[1]{} 
\newcommand{\note}[1]{}
\newcommand{\kt}[1]{} 
\newcommand{\michelle}[1]{} 
\newcommand{\noel}[1]{} 
\newcommand{\tara}[1]{} 
\newcommand{\manya}[1]{} 
\newcommand{\sunny}[1]{} 
\newcommand{\omer}[1]{} 
\newcommand{\kaitlyn}[1]{} 
\newcommand{\pgk}[1]{} 
\newcommand{\nm}[1]{} 

\newcommand{\postsubmission}[1]{#1}

\newcommand{\eg}{e.g., }

\newcommand{\ie}{i.e., }
\newcommand{\etal}{et al.\@\xspace}

\newcommand{\numpapers}{95\xspace}

\newcommand{\spas}{digital safety\xspace}

\newcommand{\spasadj}{digital-safety\xspace}

\newcommand{\cf}[1]{\factor{#1}} 
\newcommand{\factor}[1]{\emph{#1}\xspace}

\sfcode`\.=1001
\sfcode`\?=1001
\sfcode`\!=1001
\sfcode`\:=1001
\newcommand\hmm[1]{\ifnum\spacefactor=1001 \uppercase{#1}\else#1\fi}

\newcommand{\user}{\hmm{a}t-risk user\xspace}
\newcommand{\users}{\hmm{a}t-risk users\xspace}

\newcommand{\Users}{At-risk users\xspace}

\newcommand{\group}{\hmm{a}t-risk population\xspace}
\newcommand{\groups}{\hmm{a}t-risk populations\xspace}
\newcommand{\Groups}{At-risk populations\xspace}

\newcommand{\fullcirc}{\newmoon}

\newcommand{\accessperson}{access to other at-risk users\xspace}
\newcommand{\Accessperson}{Access to other at-risk users\xspace}
\newcommand{\marginalization}{marginalization\xspace}
\newcommand{\Marginalization}{Marginalization\xspace}
\newcommand{\marginalized}{marginalized\xspace}
\newcommand{\political}{legal or political\xspace}
\newcommand{\Political}{Legal or political\xspace}
\newcommand{\socialnorms}{social norms\xspace}
\newcommand{\Socialnorms}{Social norms\xspace}

\usepackage[hyphens]{url}

\usepackage{hyperref}

\renewcommand{\sectionautorefname}{Section\kern-0.2em}
\renewcommand{\subsectionautorefname}{Section\kern-0.2em}
\renewcommand{\subsubsectionautorefname}{Section\kern-0.2em}

\begin{document}
\title{{\fontsize{23pt}{\baselineskip}\selectfont SoK: A Framework for Unifying At-Risk User Research}}

\author{
\IEEEauthorblockN{
    Noel Warford\IEEEauthorrefmark{1}, Tara Matthews\IEEEauthorrefmark{2}, Kaitlyn Yang\IEEEauthorrefmark{1}, Omer Akgul\IEEEauthorrefmark{1}, Sunny Consolvo\IEEEauthorrefmark{2},\hspace{1cm}\\
    Patrick Gage Kelley\IEEEauthorrefmark{2}, Nathan Malkin\IEEEauthorrefmark{1}, Michelle L. Mazurek\IEEEauthorrefmark{1}, Manya Sleeper\IEEEauthorrefmark{2}, and Kurt Thomas\IEEEauthorrefmark{2}\hspace{1cm} }
\IEEEauthorblockA{\IEEEauthorrefmark{1}University of Maryland, \IEEEauthorrefmark{2}Google}
}
\maketitle

\begin{abstract}

\Users are people who \postsubmission{experience risk factors that augment or amplify their chances of being digitally attacked and/or suffering disproportionate harms.}
In this systematization work, we present a framework for reasoning about \users based on a wide-ranging meta-analysis of \numpapers papers. 
Across the varied populations that we examined (\eg children, activists, people with disabilities), we identified 10 unifying \emph{contextual risk factors}---such as \cf{\marginalization} and \cf{access to a sensitive resource}---that augment or amplify \spasadj risks and their resulting harms.
We also identified technical and non-technical \emph{practices} that \users adopt to attempt to protect themselves from \spasadj risks. We use this framework to discuss \emph{barriers} that limit \users' ability or willingness to take protective actions. We believe that researchers and technology creators can use our framework to identify and shape research investments to benefit \users, and to guide technology design to better support \users. 
\end{abstract}

\section{Introduction}
\label{s:intro}

Anyone can experience attacks related to their security, privacy, or safety online (\ie \emph{\spas}), but \emph{\users} have risk factors that augment or amplify their chances of being digitally attacked and/or suffering disproportionate harms. For example, some activists are surveilled by government actors due to their work~\cite{marczak2017social,sanches2020under}; people who are LGBTQ+ face elevated risk of harassment by anonymous attackers on social media~\cite{blackwell2016lgbt,blackwell2017classification}; and women in repressive regions experience pervasive sexual harassment online and sometimes severe consequences from their community as a result~\cite{vashistha2019threats, sambasivan2019they}.

A growing body of research has explored how the \spasadj needs of \users may be unmet by existing security, privacy, and safety threat models that tend to focus on a mythical ``average user.'' A common recommendation from researchers in this space is to consider \users during the technology creation process (\eg \cite{matthews2017stories,sleeper2019tough,daffalla2021defensive,barwulor2021disadvantaged,vashistha2018examining, zhao2019make}). However, for technology creators, 
it can be bewildering to consider dozens of different \groups, 
each with disjoint and sometimes contradictory \spasadj needs. Accordingly, we argue that there is a need for synthesis: to organize what is known into a framework that can be used to reason about \users' \postsubmission{risks and needs}, and to identify gaps in knowledge for future work.

We systematically identified and reviewed \numpapers papers focused on the \spasadj experiences of at-risk populations and developed a framework that can be used to reason about four research questions regarding \users:

\vspace{5pt}
\begin{enumerate}[label={\bfseries RQ\arabic*:},leftmargin=1.5cm]
    \item \emph{Contextual risk factors}. What factors---such as a person's situation in society, relationships, or personal circumstances---contribute to \spasadj risks for \users?
    \item \emph{Interactions}. How do these contextual risk factors interact to elevate the risk or severity of \spasadj attacks 
    for \users?
    \item \emph{Protective practices}. What protective practices are common across \users when attempting to address their \spasadj risks? 
    \item \emph{Barriers.} What barriers do \users encounter in protecting themselves from \spasadj risks? 
\end{enumerate}
\vspace{5pt}

Based on an analysis across 31 distinct population categories (\eg journalists, refugees, older adults), we identified 10 contextual risk factors that cross-cut \groups, yielding a set of circumstances that technology creators and researchers can consider in research, design, and development. 
We also found that \users currently rely on varied, often ad-hoc protective practices, ranging from leaning on social connections 
to relying on 
a patchwork of technical strategies to try to minimize risks and harms. We provide an \emph{at-risk framework} comprised of these \emph{contextual risk factors} and \emph{protective practices}, which we use
to discuss \emph{barriers} that limit or prevent \users from enacting digital protections, and to show how competing priorities, a lack of \spas awareness, and broken technology assumptions compound the challenges \users face.

We advocate for technology creators and researchers to consider \users' needs in risk modeling and design. Our framework provides a blueprint for addressing these issues through research, education support, and technology creation, to better ensure that \users can engage safely online, and in the process, to improve \spas for everyone. 

\section{Who are at-risk users?}
\label{s:background}

In this paper, we use \emph{\user} as an umbrella term for anyone \postsubmission{with risk factors that augment or amplify their chances of being attacked digitally and/or suffering disproportionate harms from an attack.}
We refer to groups \postsubmission{of \users}
as \emph{\groups}.  Due to a lack of consensus in the literature on how to refer to such users or populations, we chose these terms with the goal of drawing focus to external \postsubmission{risks}
these users face.

\subsection{Previous taxonomies of attacks, threats, and harms}
Previous systematizations developed frameworks to broadly categorize attacks, threats, and harms, although none capture how these elements overlap or differ across distinct \groups. In particular, Scheuerman \etal~\cite{scheuerman2021framework} and Thomas \etal~\cite{thomas2021sok} developed frameworks for understanding classes of harms that may result from \spasadj attacks, such as reputational harm, financial harm, reduced sexual safety, reduced physical safety, and coercion. Scheuerman \etal~\cite{scheuerman2021framework} also provided a framework for assessing the severity of threats based on such harms. Thomas \etal~\cite{thomas2021sok}, Sambasivan \etal~\cite{sambasivan2019they}, and Levy and Schneier~\cite{levy2020privacy} detailed how attacks vary based on the capabilities of attackers, such as having intimate access to a target, or privileged access to a target's devices or data. Our at-risk framework differs in that we isolate the contextual risk factors that can make \users particularly vulnerable to such attacks, threats, or harms. We also document common protective practices \users adopt and discuss barriers they face to staying safe.

\subsection{Value of focusing on at-risk users}
\label{ss:value}

The challenges experienced by \users can be inordinately complex, reflecting broader, societal ``structural inequalities and social norms''~\cite{mcdonald2020privacy, freed2018stalker}. These inequalities, which vary globally, mean that particular care is required to integrate \users' experiences and identities into the technology creation process~\cite{mcdonald2020privacy, keyes2019human, walker2019moving}. 

We advocate for increased focus on \users' needs by technology creators and researchers during threat modeling, research, design, \postsubmission{and development}. 
Accounting for \users 
can also elevate the \spas of all users by  
making ``more pronounced the need[s] that many of us have''~\cite{dunphy2014understanding}. 
Providing better \spasadj tools and guarantees can have far-reaching impact both to \users and general users.
Additionally, providing choices and controls for \users who know intimately the \spasadj threats they face can also benefit general users who may desire similar protections.  

\section{Methods}
\label{s:method}

We synthesize \numpapers research papers from a cross-section of computer science conferences.
Here, we discuss how we identified and analyzed these papers.
\footnote{The complete list of \numpapers papers in the dataset can be found at \textcolor{blue}{\href{https://docs.google.com/spreadsheets/u/1/d/e/2PACX-1vTy06wCGf8h4nKqU_2EB7E5b-gXKQGz0uOB6JAwHGjzBaurTbLRxEc0AcXOMIH_h5OxLTHM5-4jzg8A/pubhtml}{this link}.}}

\subsection{Paper selection}
\label{ss:selection}

Our dataset for this analysis was  \numpapers papers describing \spasadj-related issues for various at-risk populations.
We collected papers from five years (2016--2020) of conferences 
spanning the security, privacy, and human-computer interaction (HCI) communities: CCS, CHI, CSCW, IEEE S\&P, NDSS, PETS, SOUPS, and USENIX Security.
We first gathered links to every paper from these conferences on 
DBLP.\footnote{See \url{https://dblp.uni-trier.de/search}}
From those links, we collected paper titles, abstracts, and publication dates, resulting in \postsubmission{6,534} papers.

To refine this list, three researchers independently read titles and abstracts for each paper and marked them as `relevant’ to our research questions or not. At this stage, we interpreted relevance broadly, selecting any paper even slightly within scope. Papers that no researcher marked as relevant were removed. Papers marked as relevant by only one researcher were reviewed by a fourth researcher and discussed. This process identified \postsubmission{127} potentially relevant papers. 

Authors with extensive experience working with at-risk populations added 12 papers from other sources and/or from outside the target date range, in order to cover a broader range of populations, for a total of \postsubmission{139} potentially relevant papers.

\subsection{Codebook development}
\label{ss:codebook}

Our goal was to identify contextual risk factors, protective practices, and other
patterns discussed by the papers in our dataset. 
As a first step, we inductively built a codebook by analyzing, in detail, a subset of papers well-aligned with our research questions. Most of the core concepts 
in our framework were identified at this stage, although inductive refinement continued throughout our analysis. 

To select this initial subset, we extracted from the dataset an initial list of populations (e.g., survivors of intimate partner abuse~\cite{matthews2017stories}, refugees~\cite{simko2018computer}, activists~\cite{daffalla2021defensive}, children~\cite{zhao2019make}, etc.). We also synthesized an initial list of risk factors, for example, attributes of the population or the threats they faced that contributed to their \spasadj-related risks.
We then selected our subset to ensure each population and risk factor on our list was represented, 
making sure to include some papers that combined multiple risk factors (e.g., low-income African American New York City residents~\cite{elliott2015:straighttalk} and foster teens~\cite{badillo-urquiola2019:risk}). This process yielded 27 papers.

We next analyzed these 27 papers and inductively built our codebook. We used the specific population discussed in each paper\footnote{For example, Simko \etal~\cite{simko2018computer} reported on interviews with refugees \textit{and} associated caseworkers; we treated this as two populations.} as a unit of analysis~\cite{thomas2006:inductive}. One of four researchers read and summarized each paper. 
The full team used these summaries to iteratively build and refine our codebook~\cite{thomas2006:inductive}, which included categories for risk factors, protective practices, and barriers to protection. 
We met throughout the process to develop, discuss, and refine codes. These detailed summaries also enabled us to examine relationships between codes and memo early ideas on themes~\cite{braun2006:thematic}.

\subsection{Full analysis}
\label{ss:analysis}
Next, we used the codebook to analyze the remaining papers 
Because our initial paper selection steps were deliberately inclusive, we continued to refine the dataset during this phase. Any researcher could flag a paper for possible exclusion \postsubmission{if it did not address any of our research questions}; determinations were made after discussion. Our final analyzed dataset included \numpapers papers.

We randomly selected 20\% of the remaining papers, which two researchers independently coded, making minor updates to the codebook as needed. The researchers assigned codes to each paper and memoed additional relevant details and context~\cite{braun2006:thematic}. After completing this 20\% sample, the researchers calculated agreement
using Krippendorff's $\alpha$ before discussing and resolving all disagreements. The researchers then repeated this procedure on another random 20\%. Once sufficient reliability was reached, the researchers split the remaining papers between them to complete coding. 

For the contextual risk factors and protective practices, we obtained $\alpha$ = [0.88, 1.00] after the second round of coding, and for the barriers, $\alpha$ = 0.83 after the third round. As these are above the standard threshold of 0.8 for reliability~\cite{krippendorff2009}, we then proceeded to individual coding. After coding all papers, the entire research team met to review the results, identify second-level codes and themes~\cite{braun2006:thematic}, and finalize our framework.

\postsubmission{We conducted credibility checks of our framework and findings to verify they were accurate and clear. Seven experts who have worked with at-risk populations reviewed a version of this paper and met with us to discuss it. All experts found the framework sound and useful.} 

\subsection{Limitations}
\label{ss:limitations}
Our  \numpapers-paper dataset is not exhaustive of all relevant papers published in the security, privacy, or HCI communities. However, given our systematic method of compiling papers, it should reasonably represent these communities' published understanding of \users. 
Also, reflecting the current state of literature from these communities, the dataset papers skewed heavily toward Western, and specifically U.S., populations.
In addition, \postsubmission{literature from other fields, especially the social sciences, could offer relevant perspectives on the \spas of \groups. (See the appendix for details on geographic representation and researcher reflexivity.)} 

As research methods and best practices for understanding \users are still being developed---often differing from one community to the next---the papers in our dataset also often did not focus on the same issues or investigate to the same depth. 
As a result, our synthesis of the contextual risk factors, protective practices, and barriers covered in this paper may not reflect all the challenges the population in question experiences. Our coding is, instead, a reflection of the current understanding in the sampled literature on each population. 

\postsubmission{Despite these limitations, we believe this work serves as a critical first step towards recognizing contextual risk factors and protective practices that span at-risk users. We advocate for future work that builds on this framework by including broader literature and cultural perspectives.}

\section{Contextual risk factors}
\label{s:contextual}

In our meta-analysis, we identified 10 contextual risk factors that augmented or amplified \spasadj risks. These risk factors, which form the first component of our at-risk framework, include: three \factor{societal factors}, influenced by an \user's role in their society and culture;
three \factor{relationship factors} stemming from who an \user knows or interacts with; and four \factor{personal circumstances} dependent on who an \user is or their personal or professional activities. We note some common attacks for each risk factor, but do not consider being the target of a digital attack alone a risk factor. 

We capture the presence of these risk factors within the papers from our dataset in Table~\ref{tab:populations_to_factors}.
A black circle (\fullcirc) indicates that we identified at least one prior study of an \group that reported risks related to that factor. We caution that the absence of a black circle for an \group in Table~\ref{tab:populations_to_factors} does \textbf{\emph{not}} imply that the risk factor is irrelevant to the population, only that it was not reported in our dataset.

\begin{table*}[hbpt]
    \centering
    \setlength{\tabcolsep}{5pt}

\begin{tabularx}{\textwidth}{Xl|lll|lll|llll|}

\multicolumn{1}{c}{} & \multicolumn{1}{c}{} & \multicolumn{3}{|c}{\bf Societal} & \multicolumn{3}{|c}{\bf} & \multicolumn{4}{|c|}{\bf Personal} \\
\multicolumn{1}{c}{} & \multicolumn{1}{c}{} & \multicolumn{3}{|c}{\bf factors} & \multicolumn{3}{|c}{\bf Relationships} & \multicolumn{4}{|c|}{\bf circumstances} \\
\cmidrule{3-12}
\bf Population Category & 
\bf Citations & 
\rotatebox{90}{\Political} &
\rotatebox{90}{\Marginalization} &
\rotatebox{90}{\Socialnorms} &
\rotatebox{90}{Relationship with the attacker} &
\rotatebox{90}{Reliance on a third party} &
\rotatebox{90}{Access to other at-risk users} &
\rotatebox{90}{Prominence} &
\rotatebox{90}{Resource or time constrained} &
\rotatebox{90}{Underserved accessibility needs} &
\rotatebox{90}{Access to a sensitive resource} 
\\
\toprule
\rowcolor{gray!10}                                             Children & \cite{zhao2019make,mcnally2018co,ghosh2018safety,mcreynolds2017toys,kumar2019privacy,lastdrager2017effective,moser2017parents,jeong2020lime,ghosh2020circle,kumar2017no} &                                                           &                                                        &                                           \fullcirc &                                                                     &                                                        \fullcirc &                                                                     &                                                   &                                                                     &                                        \fullcirc &                                                                       \\
                                                Teens &                                    \cite{wisniewski2016dear,wisniewski2017parents,ghosh2018safety,wisniewski2017parental,ghosh2018matter,ghosh2020circle,mchugh2017most} &                                                           &                                                        &                                           \fullcirc &                                                           \fullcirc &                                                        \fullcirc &                                                                     &                                                   &                                                                     &                                        \fullcirc &                                                                       \\
\rowcolor{gray!10}                                         Foster teens &                                                                                                                                         \cite{badillo-urquiola2019:risk} &                                                           &                                                        &                                                     &                                                           \fullcirc &                                                        \fullcirc &                                                                     &                                                   &                                                                     &                                        \fullcirc &                                                                       \\
                                         Older adults &                                                                                          \cite{frik2019privacy,mcneill2017privacy,nicholson2019if,hornung2017navigating} &                                                           &                                                        &                                                     &                                                                     &                                                        \fullcirc &                                                                     &                                                   &                                                           \fullcirc &                                        \fullcirc &                                                                       \\
\midrule
\rowcolor{gray!10}                                            Activists &                                                                          \cite{daffalla2021defensive,tadic2016ict,marczak2017social,kow2016mediating,alvarado2017making} &                                                 \fullcirc &                                                        &                                                     &                                                                     &                                                                  &                                                                     &                                         \fullcirc &                                                           \fullcirc &                                                  &                                                             \fullcirc \\
                        Activists $\times$ Transgender people &                                                                                                                                                 \cite{lerner2020privacy} &                                                 \fullcirc &                                              \fullcirc &                                                     &                                                                     &                                                                  &                                                                     &                                         \fullcirc &                                                                     &                                                  &                                                             \fullcirc \\
\rowcolor{gray!10}             People involved with political campaigns $\times$ US &                                                                                                                                                   \cite{consolvo2021why} &                                                 \fullcirc &                                                        &                                                     &                                                                     &                                                                  &                                                           \fullcirc &                                         \fullcirc &                                                           \fullcirc &                                                  &                                                             \fullcirc \\
                                             Teachers &                                                                                                                                                  \cite{kumar2019privacy} &                                                           &                                                        &                                                     &                                                                     &                                                                  &                                                           \fullcirc &                                                   &                                                           \fullcirc &                                                  &                                                                       \\
\rowcolor{gray!10}                                          Journalists &                                            \cite{mcgregor2015investigating,edalatnejad2020datasharenetwork,mcgregor2016individual,mcgregor2017weakest,mcgregor2017would} &                                                 \fullcirc &                                                        &                                                     &                                                                     &                                                                  &                                                           \fullcirc &                                         \fullcirc &                                                           \fullcirc &                                                  &                                                             \fullcirc \\
                                          Sex workers &                                                                                                              \cite{barwulor2021disadvantaged,strohmayer2019technologies} &                                                 \fullcirc &                                              \fullcirc &                                                     &                                                                     &                                                                  &                                                                     &                                                   &                                                                     &                                                  &                                                                       \\
\rowcolor{gray!10}                                             ER staff &                                                                                                                                          \cite{stobert2020understanding} &                                                           &                                                        &                                                     &                                                                     &                                                                  &                                                                     &                                                   &                                                           \fullcirc &                                                  &                                                             \fullcirc \\
                                            NGO staff &                                                                                                                                  \cite{chen2019computer,le2018enforcing} &                                                 \fullcirc &                                                        &                                                     &                                                                     &                                                                  &                                                           \fullcirc &                                         \fullcirc &                                                                     &                                                  &                                                             \fullcirc \\
\rowcolor{gray!10}                                        Crowd workers &                                                                                                                                                        \cite{xia2017our} &                                                           &                                                        &                                                     &                                                           \fullcirc &                                                                  &                                                                     &                                                   &                                                           \fullcirc &                                                  &                                                                       \\
\midrule
                                        LGBTQ+ people &                                              \cite{lerner2020privacy,blackwell2016lgbt,scheuerman2018safe,hamidi2018gender,blackwell2017classification,chung2017privacy} &                                                           &                                              \fullcirc &                                                     &                                                                     &                                                                  &                                                                     &                                                   &                                                                     &                                                  &                                                                       \\
\rowcolor{gray!10}                              LGBTQ+ people $\times$ With HIV &                                                                                                                            \cite{warner2018privacy,warner2020evaluating} &                                                           &                                              \fullcirc &                                                     &                                                                     &                                                                  &                                                                     &                                                   &                                                                     &                                                  &                                                                       \\
                        Marginalized racial group $\times$ US &                                                                                                                                                        \cite{to2020they} &                                                           &                                              \fullcirc &                                                     &                                                                     &                                                                  &                                                                     &                                                   &                                                                     &                                                  &                                                                       \\
\midrule
\rowcolor{gray!10}                               People with an illness &                                                                                                \cite{petelka2020being,schmitt2018participatory,rubya2017interpretations} &                                                           &                                              \fullcirc &                                                     &                                                                     &                                                                  &                                                                     &                                                   &                                                                     &                                                  &                                                                       \\
             Older adults $\times$ With cognitive impairments &                                                               \cite{berridge2019cameras,frik2019privacy,mcdonald2020realizing,mentis2019upside,cornejo2016vulnerability} &                                                           &                                                        &                                                     &                                                                     &                                                        \fullcirc &                                                                     &                                                   &                                                           \fullcirc &                                        \fullcirc &                                                                       \\
\rowcolor{gray!10}                       People with visual impairments &                                                \cite{ahmed2015privacy,ahmed2016addressing,akter2020uncomfortable,fanelle2020blind,hayes2019cooperative,wang2019eartouch} &                                                           &                                                        &                                           \fullcirc &                                                                     &                                                        \fullcirc &                                                                     &                                                   &                                                                     &                                        \fullcirc &                                                                       \\
           People with other or multiple disabilities &                                                                                                                              \cite{marne2017learning,porter2017filtered} &                                                           &                                              \fullcirc &                                                     &                                                                     &                                                                  &                                                                     &                                                   &                                                                     &                                        \fullcirc &                                                                       \\
\midrule
\rowcolor{gray!10}                           Non-Western culture $\times$ Women &                                                                 \cite{sambasivan2019they,alghamdi2015security,vashistha2019threats,dev2020lessons,sambasivan2018privacy} &                                                           &                                              \fullcirc &                                           \fullcirc &                                                           \fullcirc &                                                        \fullcirc &                                                                     &                                                   &                                                           \fullcirc &                                                  &                                                                       \\
                     Developing regions $\times$ Older adults &                                                                                                                                                   \cite{jack2019privacy} &                                                 \fullcirc &                                                        &                                           \fullcirc &                                                                     &                                                                  &                                                                     &                                                   &                                                                     &                                        \fullcirc &                                                                       \\
\rowcolor{gray!10}                          Developing regions $\times$ Low SES &                                                     \cite{vashistha2018examining,ahmed2019everyone,ahmed2017privacy,reichel2020have,mehmood2019towards,ahmed2017digital} &                                                 \fullcirc &                                              \fullcirc &                                           \fullcirc &                                                           \fullcirc &                                                                  &                                                                     &                                                   &                                                           \fullcirc &                                                  &                                                                       \\
                           Developed regions $\times$ Low SES &                                    \cite{sleeper2019tough,vitak2018knew,redmiles2016learned,wendt2020civic,redmiles2017digital,kozubaev2019spaces,warshaw2016intuitions} &                                                           &                                              \fullcirc &                                           \fullcirc &                                                           \fullcirc &                                                                  &                                                                     &                                                   &                                                           \fullcirc &                                                  &                                                                       \\
\rowcolor{gray!10}Developed regions $\times$ Low SES $\times$ Marginalized racial group &                                                                                                                                          \cite{elliott2015:straighttalk} &                                                           &                                              \fullcirc &                                           \fullcirc &                                                                     &                                                                  &                                                                     &                                                   &                                                           \fullcirc &                                                  &                                                                       \\
\midrule
                              Undocumented immigrants &                                                                                                                                                \cite{guberek2018keeping} &                                                 \fullcirc &                                              \fullcirc &                                                     &                                                                     &                                                                  &                                                                     &                                                   &                                                                     &                                        \fullcirc &                                                                       \\
\rowcolor{gray!10}                                             Refugees &                                                                                                                                                 \cite{simko2018computer} &                                                           &                                                        &                                           \fullcirc &                                                                     &                                                        \fullcirc &                                                                     &                                                   &                                                           \fullcirc &                                        \fullcirc &                                                                       \\
                  People involved with armed conflict &                                                                                                                                                  \cite{shklovski2018use} &                                                 \fullcirc &                                                        &                                                     &                                                                     &                                                                  &                                                                     &                                                   &                                                                     &                                                  &                                                                       \\
\midrule
\rowcolor{gray!10}                          Survivors of sexual assault &                                                                                                                        \cite{obada2020towards,andalibi2016understanding} &                                                           &                                              \fullcirc &                                                     &                                                           \fullcirc &                                                        \fullcirc &                                                                     &                                                   &                                                                     &                                                  &                                                                       \\
                  Survivors of intimate partner abuse &                                                        \cite{matthews2017stories,freed2018stalker,havron2019clinical,chatterjee2018spyware,freed2019my,freed2017digital} &                                                           &                                              \fullcirc &                                                     &                                                           \fullcirc &                                                        \fullcirc &                                                                     &                                                   &                                                           \fullcirc &                                                  &                                                                       \\
\rowcolor{gray!10}                             Survivors of trafficking &                                                                                                                                                  \cite{chen2019computer} &                                                 \fullcirc &                                                        &                                                     &                                                           \fullcirc &                                                        \fullcirc &                                                                     &                                                   &                                                           \fullcirc &                                                  &                                                                       \\
\bottomrule
\end{tabularx}

    \caption{Meta-analysis showing how the contextual risk factors from our at-risk framework apply to categories of populations in our dataset. A black circle means one or more citations indicated the contextual risk factor was relevant. Note that the absence of a black circle does not mean the population does not have that \spasadj risk. Rather, it means the set of papers cited did not discuss or explore that risk factor as defined in this framework.}
    \label{tab:populations_to_factors}
\end{table*}

Next, we describe each contextual factor, focusing on the nature of the risk and types of associated attackers\footnote{We use \emph{attacker} broadly to refer to anyone who introduces digital-safety issues for an at-risk user, regardless of the severity or intention.} and harms, as applicable.
We also explore how risk factors may intersect to create more severe risks.

\subsection{Societal factors}
The first set of contextual risk factors involve \emph{societal factors} amorphously driven by cultures and institutions. 
Attacks related to societal risk factors tended to be \emph{diffusely targeted}, \ie directed toward anyone in a population or an entire \group simultaneously, rather than at a specific person.

\paragraph{\Political} 
The government, political affairs, or laws of a country can contribute to \groups experiencing heightened \spasadj risks, including potentially sophisticated attacks from government actors. A key theme associated with this factor was the power differential between government or quasi-governmental actors and the targeted populations.

Government or quasi-government actors may be able to intercept communications from \groups in various ways, such as physically seizing devices or data~\cite{mcgregor2015investigating, le2018enforcing}, 
impersonating trusted entities~\cite{marczak2017social},
coercing platform or telephony providers to bypass security measures~\cite{sanches2020under}, 
or preventing internet access entirely~\cite{daffalla2021defensive}.
For example, in 2019,
the Sudanese government shut off the country's mobile data network to make organizing for activism as difficult as possible~\cite{daffalla2021defensive}.
Similarly, the International Committee of the Red Cross---a non-governmental organization (NGO) that collects information that could be used by armed groups for non-humanitarian intelligence---reported being obligated to physically surrender devices to meet with those armed groups~\cite{le2018enforcing}.

Governments may also be able to enact surveillance, leading to real or perceived threats of monitoring. For example, undocumented immigrants in the U.S. reported concern about posting their activities on social media, due to perceived government monitoring~\cite{guberek2018keeping}. 
More generally, residents of several countries with government-controlled internet surveillance have reported modifying their behavior~\cite{vashistha2019threats, sambasivan2018privacy, jack2019privacy}.

Because of the power differential, threats associated with this factor may also escalate into offline harms, such as detainment, incarceration, or deportation~\cite{daffalla2021defensive,guberek2018keeping}. 
These harms may also have wide-ranging societal impacts, including restricted or self-censored speech~\cite{kow2016mediating, tadic2016ict} or lasting damage to trust in public institutions or figures~\cite{consolvo2021why, mcgregor2016individual}. For example, attacks on people involved with U.S. political campaigns were described as intending to undermine the institution of U.S. elections~\cite{consolvo2021why}.

\paragraph{\postsubmission{\Marginalization}}
Pervasive negative \postsubmission{treatment or exclusion} at a societal level, due to an individual's identity attributes or life experiences, may 
also elevate \spasadj risk. 

Attackers may target \groups with online hate and harassment due to their beliefs, identity, or social status, such as people who are LGBTQ+~\cite{warner2018privacy, blackwell2016lgbt,blackwell2017classification,warner2020evaluating, lerner2020privacy,scheuerman2018safe,hamidi2018gender}, undocumented immigrants~\cite{guberek2018keeping}, marginalized racial and ethnic groups~\cite{to2020they,elliott2015:straighttalk}, or people with low socioeconomic status (SES)~\cite{kozubaev2019spaces,elliott2015:straighttalk}. 
These threats were typified by a broad set of potential attackers, and contending with hate and harassment led to emotional distress~\cite{blackwell2016lgbt, to2020they, vashistha2019threats}. For this factor, past research has described the perception that anyone may be a potential attacker~\cite{scheuerman2018safe}, exacerbated by the fact that attackers may hide behind anonymous identities~\cite{to2020they, sambasivan2019they}.

Harassment experienced by \marginalized populations can be subtle or even unintentional. For example, in To et al.~\cite{to2020they}, people in marginalized racial and ethnic groups have reported experiencing widespread microaggressions pertaining to their race, via their interpersonal interactions with others on- and offline. Given the prevalence of this threat, some populations may be reluctant to fully participate online due to the risk of revealing a \marginalized characteristic~\cite{obada2020towards, andalibi2016understanding, freed2018stalker, freed2018stalker, barwulor2021disadvantaged, schmitt2018participatory, warner2018privacy, warner2020evaluating}.

Given that this factor can be associated with stigmatized information, attackers may also coerce \users by threatening to leak information that could be harmful. This can lead to reputation damage~\cite{warner2018privacy, vashistha2019threats} 
or sexual violence~\cite{guberek2018keeping, sambasivan2019they}. 
For example, attackers may threaten to ``out'' transgender individuals~\cite{lerner2020privacy} or in conservative regions, may threaten to leak chat logs between a man and a woman to damage the woman's reputation~\cite{vashistha2019threats, sambasivan2019they}. 
Algorithmic bias may also create or exacerbate \spasadj risks in this space. For example, automated gender recognition systems may misgender transgender people~\cite{hamidi2018gender}.

\paragraph{\postsubmission{\Socialnorms}}
\Socialnorms are informal rules that govern behavior in society. Some norms can particularly restrict options for members of an \group, leading to increased \spasadj risks when there is a mismatch between the expectations of technology creators and the lived experiences of that population.

For example, technology creators may assume that devices and accounts are personal and private. However, device-sharing norms mean that this assumption may not hold for a variety of \groups, limiting the privacy protections afforded by private personal devices and accounts. For example, Sambasivan et al. \postsubmission{and Ahmed et al.} found that women in South Asia were expected to share devices or accounts with family members~\cite{sambasivan2018privacy,ahmed2017digital}. Similar device sharing norms were also found among rural women in Greenland~\cite{wendt2020civic}. 

Technology creators may also assume that users understand implicit norms around when it is appropriate (or dangerous) to share personal information. However, some \groups face changing norms, which can lead to unexpected threats. For example, alongside numerous other challenges, refugees must adapt to the technology norms and associated \spasadj risks of their new country of residence. For refugees in the U.S. specifically, certain types of information, such as Social Security numbers, were more sensitive than expected, placing refugees at increased risk of financial harm until they learned when it was appropriate to disclose this type of information~\cite{simko2018computer}.

\subsection{Relationships}
The second set of risk factors is driven by the \emph{relationships} of \users, including direct relationships with an attacker and relationships with a third party. These risks tended to be associated with \emph{focused targeting}, in which attackers pursue specific \users (often with intense motivation).

\paragraph{Relationship with the attacker} 
A personal relationship with an attacker can lead to heightened risk of what Levy and Schneier~\cite{levy2020privacy} classify as ``intimate threats.'' It can be particularly difficult for individuals targeted by intimate threats to prevent and detect harm, because the attackers are likely to have physical access to the target's devices and accounts, implicit or explicit authority over the target, and potentially detailed knowledge about the target they can leverage~\cite{levy2020privacy}. Intimate threats can involve a wide range of \spasadj attacks including surveillance, device or account compromise, destruction of data or devices, harassment, and more~\cite{levy2020privacy}.

For example, survivors of intimate partner abuse~\cite{matthews2017stories, freed2018stalker, havron2019clinical, freed2019my} were reported as often facing relentless attacks from abusers aiming to limit the survivor's autonomy. These abusers may have physical or digital access to the survivors' devices and accounts, including surveillance via coercive physical access~\cite{matthews2017stories} or spyware~\cite{freed2018stalker, chatterjee2018spyware}. Similarly, survivors of trafficking reported worrying about being recaptured because, prior to escape, the trafficker had full access to the survivor's digital life~\cite{chen2019computer}. Stalking attacks are also often 
characterized by a \cf{relationship with the attacker}, as ``the majority of stalkers are \ldots obsessionally focused on a specific person with whom they have had some previous relationship,'' often an intimate one~\cite{davis2000research}.
\postsubmission{Workers on online crowdsourcing platforms also experienced a severe power imbalance with requesters on those platforms, who could decide whether or not to pay workers based on their responses, and could gather detailed information on them via completed tasks~\cite{xia2017our}.}

\paragraph{Reliance on a third party} Individuals may be at risk due to their \cf{reliance on a third party} for safety-focused care or help with essential tasks.  Unlike intimate threats associated with a \emph{relationship with the attacker}, the third parties in this context typically have helpful or safety-focused reasons for ``privacy invasions''~\cite{levy2020privacy}. However, even with generally supportive intentions, these invasions can increase the individual's \spasadj risks due to reduced privacy (from the third party), leading \users to feel uncomfortable or that their autonomy is limited~\cite{mentis2019upside}. These types of risks can also increase the attack surface (e.g., through the third-party), or leave \users vulnerable to attackers that impersonate the third party.

For example, children and teens often shared their devices and personal data with parents or caregivers~\cite{mcnally2018co,kumar2017no}, or had applications or privacy settings enabled that allowed parental monitoring~\cite{ghosh2020circle, wisniewski2017parental, ghosh2018matter}. 
People with visual impairments sent photos to crowdsourced services for object identification, but also worried that there might be sensitive content in the photos (e.g., credit card numbers)~\cite{akter2020uncomfortable}. Older adults with cognitive impairments depended on help from caregivers for digital tasks~\cite{frik2019privacy,mentis2019upside}. Refugees relied on case workers for processing personal data, applying to jobs, or applying for support~\cite{simko2018computer}.

\paragraph{\Accessperson}
Having access to another at-risk person or population \postsubmission{(the ``primary target'')} can also put someone at greater risk of focused, stepping-stone attacks that ultimately aim to harm the \postsubmission{primary target}. For example, children of survivors of intimate partner abuse may be targeted by abusers in order to regain access to the survivor~\cite{matthews2017stories}, journalists may be targeted to try to gain access to their sources~\cite{mcgregor2017weakest, mcgregor2016individual, mcgregor2015investigating, edalatnejad2020datasharenetwork,mcgregor2017would}, and elementary school teachers may be targeted to gain access to their students~\cite{kumar2019privacy}.

\subsection{Personal circumstances}
\label{ss:circumstances}
Our third set of contextual risk factors cover \emph{personal circumstances} that can increase risk, such as public prominence or socioeconomic constraints. All of these factors tended to involve \emph{diffuse targeting} of groups, except \cf{prominence}, which involved \emph{focused targeting} of specific individuals. 

\paragraph{Prominence} 
\Users who stand out in a population, because they are well-known publicly or have noticeable attributes (e.g., accomplishments, outspokenness, attractiveness, etc.), may face heightened risk associated with their prominence. \cf{Prominence} can expose individuals to new attacks, like \emph{focused targeting}. Celebrities, for example, may experience parasocial relationships with audiences, increasing the risk of stalking or personal information leaking~\cite{celebrity-parasocial}. 

\cf{Prominence} may also exacerbate risks associated with \cf{\political} factors or \cf{\marginalization}, a dynamic identified for prominent journalists~\cite{mcgregor2015investigating, edalatnejad2020datasharenetwork, mcgregor2016individual, mcgregor2017weakest}, politicians~\cite{consolvo2021why}, NGO staff~\cite{le2018enforcing}, and activists~\cite{daffalla2021defensive, marczak2017social}. In these cases, the level of prominence influenced the severity of risks. For example, people involved in a national U.S. political campaign were more at risk of focused attacks than campaigns receiving only state or local attention~\cite{consolvo2021why}.

\paragraph{Resource or time constrained}
Populations can experience elevated \spasadj risks and decreased ability to respond to \spasadj threats if they have limited access to technology or other resources (e.g., money, devices, connectivity), or limited ability to set aside time to cope with risks.
\Users experiencing this risk factor face constraints that go beyond what is typical. Individuals with low-SES, for example, may not be able to afford private personal devices, or may need to rely on old devices that can no longer receive security updates~\cite{elliott2015:straighttalk,vashistha2018examining, sleeper2019tough, vitak2018knew}.

\Users can also face extraordinary time constraints. 
Political workers, for example, operated within extremely limited election timelines, which made it challenging to set up the security infrastructure needed to counter nation-state attackers~\cite{consolvo2021why}. 
Hospital emergency departments similarly ``have strong availability demands ... and must provide services as quickly as possible''~\cite{stobert2020understanding}, leading to password reuse and use of unsecured personal devices in a sensitive work context.

\paragraph{Underserved accessibility needs}
Some \users have accessibility needs that are underserved by current technology, contributing to their \spasadj risks. In our dataset, this included accessibility needs due to a disability, neurodiversity, a language barrier, or developmental maturity. Inaccessible technology can cause anxiety about, and susceptibility to, potential attacks.

Members of populations experiencing this risk factor 
have described general anxiety about falling prey to ``hackers'' or vague bad actors, as well as worries about their ability to effectively protect themselves with existing technology. For example, some older adults reported asking trusted sources for \spasadj help because they did not feel confident protecting themselves~\cite{frik2019privacy}. This can stem from negative past experiences or from inaccessible  language in online resources about \spas~\cite{nicholson2019if}. Assumptions built into systems were unrealistic for older adults with mild cognitive impairments, who were sometimes unable to remember passwords and other crucial information~\cite{mentis2019upside}. In some cases, they had trouble remembering whether they made a particular purchase~\cite{mcdonald2020realizing}, making it difficult to differentiate an attack from a memory lapse.

Similarly, people with disabilities may struggle with a lack of accessible technology for some tasks~\cite{ahmed2015privacy,ahmed2016addressing,akter2020uncomfortable,fanelle2020blind,hayes2019cooperative,wang2019eartouch,marne2017learning}. For example, people with visual impairments reported to Wang \etal the conflict between the need to use screen readers in public and concerns about eavesdropping and safety~\cite{wang2019eartouch}.

\postsubmission{Additionally, r}efugees with developing English skills reported struggling with language accessibility, finding it difficult to distinguish legitimate callers from scammers when the call was in English~\cite{simko2018computer}. 
Zhao \etal found that children do not completely understand certain online privacy risks due to their age and development~\cite{zhao2019make}.

\paragraph{Access to a sensitive resource} 
Access to a sensitive resource (\eg sensitive data, credentials, money) can increase the risk of attacks aimed at co-opting this access. In most cases, the \users' professional activities provided them privileged access to these resources.

For example, emergency department staff may be targeted for their access to patient medical data~\cite{stobert2020understanding}, journalists for their access to original source material, such as legal documents or financial records~\cite{mcgregor2016individual, edalatnejad2020datasharenetwork, mcgregor2015investigating}, and executive staff for their ability to approve wire transfers~\cite{cfo-attack}.

\subsection{How do contextual risk factors interact?}
\label{ss:factor_combinations}

Most \groups in Table~\ref{tab:populations_to_factors} experienced more than one contextual risk factor. Each factor contributed to risk on its own, but risk factors also combined to yield new \spasadj risks or amplify existing risks. Thus, we argue that technology creators and researchers should consider all of an \group's risk factors together when possible. This is related to prior work on intersectionality~\cite{schlesinger2017intersectional,crenshaw1989demarginalizing}, which considers how multiple marginalized identities or circumstances can combine to create unique modes of discrimination. We provide an illustrative list of examples from our thematic analysis below, \postsubmission{chosen because they represented experiences reported in multiple papers across populations and/or provided understandable illustrations of how risk factors could combine to yield new risks or amplify each other.}

\paragraph{\cf{Prominence} added \emph{focused targeting} to other risk factors}
When added to other risk factors, \emph{prominence} appeared to make it more likely that an at-risk user would experience \emph{focused targeting} that amplified their other risk factors. For example, political campaign workers had \emph{access to a sensitive resource}, but \emph{prominent} politicians were more likely to be subjected to \emph{focused targeting} to access those sensitive resources~\cite{consolvo2021why}. Similarly, while many transgender people reported experiencing \emph{\marginalization}, the \emph{prominence} of transgender activists resulted in highly targeted hate and harassment attacks~\cite{lerner2020privacy}.

\paragraph{\cf{Resource or time constraints} made it harder to cope with other risk factors}
Populations who are \cf{resource or time constrained} were more likely to experience worse outcomes pertaining to their other risk factors, because they did not have the time, money, or other resources to effectively protect themselves or recover from attacks. 
For example, the primary risk factor for survivors of intimate partner abuse was typically their \cf{relationship with the attacker}, but prior research has emphasized that survivors with low SES had particular difficulty protecting their (and their children's) digital lives~\cite{matthews2017stories}.

\paragraph{\cf{Underserved accessibility needs} and \cf{reliance on a third party} often combined}
In our dataset, \users with \cf{underserved accessibility needs} also tended to \cf{rely on a third party} for care or help with technology, at least sometimes. This included children, teens, older adults, people with visual impairments, and refugees, as shown in \autoref{tab:populations_to_factors}. For example, older adults experiencing mild cognitive impairment often forgot important passwords and information (related to \cf{underserved accessibility needs}), which reinforced the need to share these passwords and information with caregivers (subjecting them to risks associated with \emph{reliance on a third party})~\cite{mentis2019upside}.

\section{Protective Practices}
\label{s:protective}

\postsubmission{\Users employed practices they perceived would help them prevent, mitigate, or respond to \spasadj risks. These \emph{protective practices} form the second part of our at-risk framework. The risk factors presented above helped drive user decisions about which protective practices to use, but these practices were not always ideal or even effective. Protective practices involved tradeoffs and highlight barriers to technology use (which we explore in \autoref{s:risky}). Different users weighed the pros and cons differently, sometimes leading to seemingly contradictory choices. We catalog these imperfect practices to show what \users currently do given their risks, and to set the stage for a discussion of barriers to protections in \autoref{s:risky}, both of which provide context for how to design technologies intended to support \users.}

Our meta-analysis identified three categories of protective practices: \emph{social strategies} where \users relied on their social connections to respond to threats; \emph{distancing behaviors} where \users distanced themselves from, or entirely abandoned, certain accounts and technologies; and \emph{technical solutions} that involved leveraging technical tools and mechanisms to prevent or respond to threats. We found these strategies were not mutually exclusive, with \users commonly relying on multiple strategies simultaneously.

\subsection{Social strategies}
\label{ss:social}
\Users frequently relied on social connections (in-person or online) to overcome \spasadj threats. This included relying on family or peers for trusted advice and support, vetting the identities of people they interacted with online, and controlling social interactions to minimize harms.

\paragraph{Informal help from trusted family and peers}
A popular protective practice among \users experiencing the \cf{\marginalization}, \cf{\socialnorms}, or \cf{underserved accessibility needs} factors was to informally seek help from trusted family and peers. The \user may be seeking direct help from someone perceived to be more knowledgeable or resourced, or may simply be seeking emotional support. For example, children and teenagers reportedly sought help from parents to understand security warnings~\cite{jeong2020lime} or deal with strange, scary, or confusing internet
experiences~\cite{zhao2019make,wisniewski2016dear,wisniewski2017parents}. Older adults who self-identified as having low technical understanding sought assistance from family members when addressing \spasadj concerns~\cite{frik2019privacy,nicholson2019if}. At times, women in South Asia relied on family members for emotional support when harassed online~\cite{sambasivan2019they, sambasivan2018privacy}.
Informal support sometimes also came from anonymous peers, such as transgender individuals relying on social media to connect with other LGBTQ+ peers for emotional support when they experienced hate and harassment~\cite{scheuerman2018safe, lerner2020privacy}, or people in armed conflict zones connecting to share critical information~\cite{shklovski2018use}.

\paragraph{Formal help from trusted organizations} Public recognition of \spasadj threats facing \users---particularly with respect to the \cf{\political}, \cf{\marginalization}, \cf{relationship with the attacker}, or \cf{resource or time constrained} risk factors---has prompted trusted organizations to offer support. For example, organizations that assist survivors of trafficking and intimate partner abuse, including NGOs~\cite{chen2019computer} and government agencies~\cite{freed2018stalker,freed2019my,havron2019clinical}, have provided assistance setting up \spasadj protections that are interwoven with continuous care~\cite{havron2019clinical}.
Similarly, NGOs have assisted refugees with technology-required tasks, such as applying for work or submitting legal documents~\cite{simko2018computer}. Public institutions, like libraries, have provided assistance and access to computers and connectivity to people with low SES~\cite{vitak2018knew}. While \users may rely on these institutions for a broad variety of technical assistance~\cite{simko2018computer,vitak2018knew}, \spasadj support was often explicitly reported as a key benefit.
This assistance could incur additional \spasadj risks due to \cf{reliance on a third party}, but the information and aid provided was often critical, both for \spas and other basic needs.

Formal support may also stem from dedicated professional resources, particularly for those who have \cf{access to a sensitive resource} or \cf{\accessperson}.
For example, journalists have benefited from institutional support and accumulated best practices, as exemplified by the International Consortium of Investigative Journalists' (ICIJ) tools for \spas in the Panama Papers investigation~\cite{mcgregor2017weakest}. 
Elementary school teachers using technology in classrooms---who have a responsibility for ensuring the \spas of their students---have sought assistance from school media specialists or librarians~\cite{kumar2019privacy}. 
Political campaign workers in the U.S. have received training from organizations such as Defending Digital Campaigns and the Center for Democracy \& Technology~\cite{consolvo2021why}.

\paragraph{Vetting identities to avoid potential attackers}
\Users who experience \cf{\marginalization} or \cf{\political} risk factors have a documented need to verify that new people they encounter online are safe to associate with or to admit into a private space. For example, social media community moderators for transgender communities of color asked potential new members questions about topics relating to race and LGBTQ+ issues to help protect their spaces from potential attackers~\cite{lerner2020privacy}. 
Women from non-Western cultures scrutinized online profiles to prevent men from infiltrating their private discussion spaces and subjecting them to sexual harassment~\cite{sambasivan2019they}. 
Sex workers informally shared lists of abusive clients and aggressors to be avoided~\cite{barwulor2021disadvantaged}. 
Undocumented immigrants restricted communication with untrusted parties until sufficiently vetted, due to concerns of impersonation by the police or immigration enforcement~\cite{guberek2018keeping}. 
\tara{@Noel: This example is vague -- how were untrusted parties vetted? Impersonation of who? Can you clarify?} \noel{I don't think I have space here to explain it with sufficient nuance, especially in the context of this list of examples. I think people who want to know more could consult the paper}
Activists used physical in-person meetings or mutually trusted peers to vet new individuals before adding them to sensitive group chats~\cite{daffalla2021defensive}. 
Unifying these practices is a lack of dependable digital signals of authenticity or identity, resulting in \users relying on ad-hoc techniques to establish trust.

\paragraph{Preemptive disclosure for control}
\Users experiencing the \cf{\marginalization} or \cf{\political} risk factors sometimes proactively revealed sensitive personal information in a controlled manner to disempower attackers and preempt future emotional harm. 
For example, some men seeking men on dating apps opted to publicly disclose their HIV-positive status, both to simplify navigating sexual negotiations and as a path toward destigmatization~\cite{warner2018privacy,warner2020evaluating}. 
Some transgender people chose to ``out'' themselves to avoid the emotional stress of maintaining a secret identity and the risk of coercion or extortion~\cite{lerner2020privacy}.
\postsubmission{Some people with disabilities---especially those with ``visible'' disabilities---chose to disclose these disabilities on dating apps, to reduce the chance of undesirable connections~\cite{porter2017filtered}.}
Some activists engaged in an ``anti-surveillance tactic of openness'' by collapsing public and private personas, rendering information they shared---like political beliefs---unusable against them in a separate context if leaked~\cite{sanches2020under}. 
This practice is not applicable to all \users, as it requires a willingness to be hyper-public, which carries potential risks akin to those that \cf{prominent} \users face.

\paragraph{Social pleas}
Some \users who experience the \cf{\marginalization} or \cf{\socialnorms} risk factors reported reaching out directly to attackers with a plea to cease activities, such as content leaks or harassment. For example, some women in South Asia either independently, or with the assistance of family, asked attackers within their community to cease online sexual harassment~\cite{sambasivan2019they, vashistha2019threats}. Similarly, some teens engaged with peers~\cite{wisniewski2016dear} or parents~\cite{moser2017parents} to have embarrassing content removed from social media. This practice leverages the attacker's empathy to help resolve \spasadj threats.

\subsection{Distancing behaviors}
\label{ss:distancing}
The second category of protective practices in our framework involves \emph{distancing}, or limiting use of technology. Distancing behaviors were prevalent in our dataset, used by most populations. The fact that limited or non-participation often felt like the safer choice may perhaps be a troubling signal to technology creators. 
Here we highlight two common themes.

\paragraph{Censoring online sharing}
Some \users carefully self-censored personal content they shared online. This was especially common for users experiencing the~\cf{\marginalization} risk factor, who often did not feel safe sharing personal information or did not want to reveal stigmatized aspects of their lives to broad online audiences. For example, some LGBTQ+ parents refrained from sharing family or personal photos to try to avoid ``outing'' themselves beyond specific social circles~\cite{blackwell2016lgbt}. In another case, some HIV-positive men seeking men reported not sharing their HIV status on dating apps to avoid unwanted stigmatization~\cite{warner2018privacy}. (As noted in the previous section, some members of this population did preemptively disclose their HIV status, demonstrating that protective practices within a population can vary.)

\paragraph{Reducing one's digital footprint}
Some \users dramatically reduced technology use to avoid attackers with extensive access, knowledge, and power, as was often the case with the ~\textit{relationship with the attacker} and ~\textit{\political} risk factors.
For example, Chen \etal~\cite{chen2019computer} found that survivors of human trafficking took relatively extreme measures to try to protect their new location, including abandoning devices, deleting social media accounts, and severely restricting online sharing. 
Survivors of intimate partner abuse employed similar measures when they suspected devices or accounts may have been compromised by their abuser~\cite{freed2017digital}.
As another example, activists in the Sudanese Revolution, concerned about the government confiscating their devices, used coded communications and regularly deleted sensitive data~\cite{daffalla2021defensive}.

\subsection{Technical solutions}
\label{ss:technical_solutions}

\Users also employed technical solutions to protect their \spas, such as specialized software or settings in common apps and services. Across these strategies, \users did their best to protect themselves based on their knowledge 
and experience; however, the technical solutions they chose did not always provide the desired protections.

\paragraph{Secure communication and encryption}
\Users experiencing the \cf{\political} risk factor, such as journalists, activists, or politicians, commonly reported using encrypted messaging platforms. 
These \users frequently wished to secure their communications against monitoring by a nation-state (or similarly resourced) attacker. 
For example, journalists working together to report on the Panama Papers used PGP 
as part of organizational security policies established by the ICIJ~\cite{mcgregor2017weakest}. Similarly, journalists 
and activists in varied international contexts reported using encrypted chat apps 
to communicate over networks directly controlled by their nation-state 
adversaries~\cite{daffalla2021defensive,sanches2020under}.
Transgender activists in the U.S., particularly those engaged in more \cf{prominent} activities, similarly used encrypted chat to protect their communications~\cite{lerner2020privacy}. NGOs, who had \cf{\accessperson}, sought to protect those \users by using encrypted email for internal communications~\cite{chen2019computer}.

Beyond encrypted messaging, some \users experiencing the \cf{\political} risk factor reported encrypting files or entire devices to protect content from attackers---including nation-states and other focused attackers---who might gain physical access. \Users employing this technique included journalists, human rights organizations, and activists~\cite{mcgregor2015investigating, marczak2017social, le2018enforcing}. 

\paragraph{Strong(er) authentication}
Strong authentication is commonly advised for internet  users in general~\cite{redmiles2020comprehensive, nist800-63}, but can be especially important for users at higher risk of device or account compromises. While truly strong authentication practices were rare in our dataset, we identified a few authentication trends in subsets of select populations. 

While frequently changing passwords is no longer considered a best practice generally~\cite{nist800-63}, some \users did this to address specific practical concerns. For example, \users who have a \cf{relationship with the attacker}, such as survivors of intimate partner abuse or trafficking, coped with attackers who may have ready access to their passwords (e.g., through coercion, co-presence leaks, or remote surveillance~\cite{freed2019my, matthews2017stories}) by regularly changing passwords on  accounts known to the attacker~\cite{freed2019my, chen2019computer}. Some users with visual impairments also chose to change passwords frequently in case others might have observed them inputting their passwords~\cite{ahmed2015privacy}.

While not mentioned frequently in our dataset, our analysis revealed some cases of successful adoption of two-factor authentication (2FA). 
For example, in Matthews et al.~\cite{matthews2017stories}, only a small number of survivors of intimate partner abuse in their study reported using 2FA to protect their accounts, despite highly motivated attackers. Only a minority of participants involved with political campaigns reported adopting the strongest form of 2FA, despite commonly experiencing phishing attacks~\cite{consolvo2021why}. Journalists collaborating on the Panama Papers (\cf{access to a sensitive resource}) were required by the ICIJ to use 2FA~\cite{mcgregor2017weakest}, though journalists are more broadly described as having limited awareness of 2FA's benefits~\cite{mcgregor2016individual}.\footnote{We note that 2FA has become more commonplace since some of these papers were published.}

\paragraph{Privacy settings and access control} We identified three categories of privacy settings (within widely used apps and services) that were discussed by multiple papers across a variety of \groups and risk factors: location privacy~\cite{frik2019privacy, dev2020lessons, matthews2017stories, chen2019computer}, social media visibility settings~\cite{jack2019privacy, dev2020lessons, chen2019computer, zhao2019make, frik2019privacy, matthews2017stories}, and blocking undesired contacts~\cite{sambasivan2019they, reichel2020have, dev2020lessons, matthews2017stories, frik2019privacy}. For users with a \cf{relationship with the attacker} or \cf{\accessperson}, these privacy settings were commonly discussed as protecting highly sensitive information that, if obtained by an attacker, could compromise safety~\cite{freed2018stalker, matthews2017stories, chen2019computer}. 
For example, some women in South Asia used platform controls to block unwanted contact from abusers on social media~\cite{sambasivan2019they}.

For \users in professional settings with \cf{access to a sensitive resource}---such as NGO staff, political campaign workers, and journalists---access control settings were also important for limiting who could access sensitive digital resources~\cite{le2018enforcing, consolvo2021why, mcgregor2017weakest}.

\paragraph{Online identity management} 
Some \users reported attempting to hide their identity online 
via careful account management, including multiple and pseudonymous accounts.

Some \users used multiple accounts or devices to maintain boundaries between different facets of their identities. \cf{\Marginalization} was one driver for this behavior; for example, sex workers (whose profession was stigmatized) reported keeping separate work and personal accounts~\cite{barwulor2021disadvantaged}.
Alternatively, survivors of sexual assault used throwaway accounts to seek support online without revealing their identities~\cite{andalibi2016understanding}.

\Users also used multiple accounts or devices in response to potential account hijacking or surveillance. Survivors of intimate partner abuse, whose \cf{relationship with the attacker} tended to give the attacker opportunities to access their accounts or devices, reported creating new accounts or purchasing new devices to avoid revictimization after escape~\cite{matthews2017stories}. Users experiencing \cf{\political} risks, such as activists, 
reported using multiple accounts and devices to protect themselves from potential nation-state attackers. For example, activists in the Sudanese Revolution reported using 
SIM cards from other countries, creating fake U.S. phone numbers 
online, and asking relatives and friends overseas to verify social media 
accounts in an effort to thwart government surveillance~\cite{daffalla2021defensive}. 
Marczak \etal found similar evidence of an activist using multiple SIM cards 
to try to prevent governments from linking calls made in different countries 
and contexts~\cite{marczak2014governments}.

In some cases---particularly cases of \cf{\marginalization}---rather 
than use entirely new accounts, \users reported 
ad-hoc pseudonymity strategies for existing accounts. 
For example, in order to prevent personal photos being used for digital abuse, some women in South Asia chose neutral images, like flowers, as profile photos~\cite{sambasivan2019they}.
This strategy was shared by some transgender people, particularly activists (who also experienced elevated risk due to \cf{prominence})~\cite{lerner2020privacy}. Low-income Black Americans similarly reported sometimes using emojis rather than contact names to protect their contacts' identities on devices they thought might be compromised~\cite{elliott2015:straighttalk}.

\paragraph{Network security}  Only two \groups in our review were reported to have used network security tools---like Tor or virtual private networks---to disguise their web traffic. Certain activists~\cite{daffalla2021defensive, marczak2017social} and journalists~\cite{mcgregor2015investigating} were aware of and used these technologies, but these tools did not appear elsewhere in our dataset---and not every user in those populations found these tools equally useful or usable~\cite{mcgregor2016individual, marczak2017social}. This could mean that other \groups are largely unaware of these tools, that existing tools are not perceived as 
useful for the particular \spasadj concerns of other \groups, or simply that 
researchers have not fully investigated this question with other \groups.

\paragraph{Tracking and monitoring applications}
Caregivers (of \users who \cf{rely on a third party}) sometimes used tracking or monitoring applications with the goal of protecting the at-risk user's \spas.
Parents of foster teens reported using router settings to limit internet access at certain times of day and parental control apps to watch for potentially dangerous behavior~\cite{badillo-urquiola2019:risk}. These applications can create conflict between autonomy and privacy, but some children and teenagers 
indicated that they can be helpful, particularly when used as part of an 
ongoing dialogue with parents~\cite{ghosh2018safety,ghosh2020circle}.

\section{Barriers to Protective Practices} 
\label{s:risky}
In the previous section, we documented a range of protective practices \users employed, motivated by the risk factors they experienced. In our thematic analysis, we also identified a variety of barriers that limited or prevented \users from \postsubmission{effectively} adopting these practices. We apply the contextual risk factors and protective practices of our at-risk framework to discuss three categories of barriers: \emph{competing priorities}, \emph{lack of knowledge or experience}, and \emph{broken technology assumptions}.

\subsection{Competing priorities}
\label{ss:deliberate}

Digital safety is not and cannot always be the top priority for all 
users; this is perhaps even more true for at-risk users with competing, often critical, needs. Some of the many competing priorities that appeared in our dataset included basic needs like food, income, or physical health and safety; social participation and compliance with social norms; and caring for others. We found that specific competing priorities were often associated with particular risk factors. 

It is well understood that in general, users prioritize convenience, simplicity, and their tasks and goals over \spas~~\cite{herley2009so,adams1999users}. Our analysis revealed a tendency for \users to have layered or more severe conflicts between \spas and other needs,
making the leap to safer behaviors especially difficult.

\paragraph{Basic needs} 
People who are \cf{resource or time constrained} in our dataset often prioritized other critical needs over \spas, despite the potential for increased \spasadj risk. For example, Elliott and Brody~\cite{elliott2015:straighttalk} reported that low-income Black New Yorkers used apps to find cheaper food, despite suspecting the apps were insecure. Similarly, people experiencing homelessness described discomfort using public Wi-Fi, but used it anyway for critical needs such as applying for government assistance, housing, and jobs~\cite{sleeper2019tough}. 

Relatedly, people who \cf{rely on a third party} often gave up control of \spas to accomplish basic tasks. Refugees, for example, shared account information, including passwords, with caseworkers to obtain social services or apply for jobs~\cite{simko2018computer}. 
Some people who are visually impaired used crowdsourced assistive technology for tasks like identifying medicines correctly, despite the risk of exposing sensitive content in the 
background of the photos crowdworkers would evaluate~\cite{akter2020uncomfortable}. 
Frik et al.\ \cite{frik2019privacy} found that some older adults were willing to accept in-home surveillance, giving up privacy to maintain some autonomy and independent living.

\paragraph{Participation and connection}
As noted in \autoref{ss:distancing} above, 
people who experience \cf{\marginalization} sometimes chose to distance or fully
disconnect in order to keep themselves safe. Members 
of these populations who opted to engage were often reported as knowing that it 
increased their risk. For example, Blackwell et al.\ reported on LGBT+ parents 
implicitly ``outing'' themselves via ``everyday'' social media posts about 
their children and families~\cite{blackwell2016lgbt}. 

We noted similar behavior related to \cf{\political} risk, where the need 
to communicate with other activists, both to organize events and simply to 
be part of a community, motivated potentially risky modes of communication~\cite{sanches2020under}. 

Similarly, maintaining good social standing in a family or community 
sometimes required people experiencing the \cf{\socialnorms} factor to deprioritize \spas. Our dataset included, for 
example, women in South Asia and Saudi Arabia, as well as people in South Africa, 
sharing accounts, devices, and credentials with family members to meet cultural 
or community expectations~\cite{alghamdi2015security,sambasivan2019they,sambasivan2018privacy, reichel2020have}. 
Some users in South Africa accepted Facebook friend requests 
from strangers, even though they were uncomfortable with the resultant potential for 
sharing, to avoid being rude~\cite{reichel2020have}. Teenagers noted that they did not discuss or share risky \spasadj experiences with their parents to avoid awkwardness or a possible negative reaction \cite{wisniewski2017parents}.

\paragraph{Caring for others}
Another example of competing priorities involved \users taking on 
additional risk to help, care for, or support others. We 
noted this occurring frequently in connection with the \textit{\marginalization} factor, in part to reduce overall stigma via normalization. 
For example, Warner et al.\ found that some gay men disclosed their 
HIV+ status on a dating app in part to increase visibility~\cite{warner2018privacy}; 
similarly, separate studies found that LGBTQ+ parents and transgender individuals 
posted on social media, despite possible risks, in part to increase visibility 
and reassure others that they were not alone~\cite{blackwell2016lgbt,lerner2020privacy}. 

Our analysis revealed cases in which \users prioritized others' autonomy over their own \spas, particularly in the case of \cf{\accessperson}. 
McGregor et al.\ ~\cite{mcgregor2016individual} quoted a journalist who valued avoiding ``impos[ing] any kind of burden 
on a source'' over the journalist's own \spas.
Similarly, NGO staff with \cf{\accessperson} reported the need to balance security with the autonomy of the survivors they worked with: Chen et al.\ ~\cite{chen2019computer} quoted a staff 
member working with trafficking survivors who tended to suggest that they turn off phone location, 
``But I do kind of leave it up to them. It’s not mandatory\ldots. ’Cause we come from an
empowering place. We don’t want to be telling
them what to do.''

Caring for others can also lead at-risk users to prioritize efficiency 
in achieving their primary goals over \spas, something we noted particularly in the cases of \cf{\political}, \cf{prominence}, or \cf{access to a sensitive resource}, which frequently coincided with a profession or activity that benefited others. 
A human rights activist, for example, wondered, ``Should I spend half a day figuring out digital security, or do work?''~\cite{marczak2017social}. 
In several studies, journalists, political campaign workers, and emergency department personnel reported mixing personal and work data and devices for efficiency in their work~\cite{mcgregor2016individual, sanches2020under, stobert2020understanding, consolvo2021why}.

\subsection{Lack of knowledge or experience}
In our dataset, nearly all at-risk populations had less \spasadj knowledge than they needed to mitigate the risks they faced. 
While prior work has found that general users tend to have limited \spasadj knowledge~\cite{wash2015too}, this was exacerbated for at-risk users who, by definition, faced serious \spasadj risks and thus usually needed to understand and deploy more robust protections than a typical user. 
The at-risk framework helps us understand current patterns in how \users experience \spasadj knowledge limitations.

\paragraph{\cf{\Political} and \cf{prominence} tended to attract sophisticated attackers, leading to stark knowledge asymmetries} It was difficult for \groups---like activists, undocumented immigrants, journalists, or political campaign workers---to surmount the knowledge differential needed to counter the nation-state attackers that might or did target them~\cite{consolvo2021why, daffalla2021defensive,tadic2016ict,mcgregor2016individual, mcgregor2015investigating}. For example, undocumented immigrants in the U.S. were often unaware that Immigration and Customs Enforcement might target them for surveillance via government requests for information to the social media platforms they used~\cite{guberek2018keeping}.

\paragraph{\cf{Relationship with the attacker} involved intimate threats that required expertise to counter} Even in the absence of sophisticated attacks, these threats can require substantial, robust protective measures. For example, survivors of intimate partner abuse contended with a motivated attacker who may launch repeated attacks leveraging intimate knowledge about them, physical access to their devices and accounts, and relational power~\cite{consolvo2021why, freed2017digital, levy2020privacy}. 

\paragraph{Some \cf{resource or time constrained} users had limited technology experience} Examples included some people with low SES, people living in developing regions, and survivors of intimate partner abuse or trafficking, who did not have regular access to new or trusted devices or the internet, which limited their ability to gain technology experience and skills~\cite{sleeper2019tough, jack2019privacy, chen2019computer, matthews2017stories}. Other populations, like journalists and people involved with political campaigns, commonly did not have the time to develop the technical skills to counter the \spasadj risks they faced~\cite{mcgregor2015investigating, mcgregor2016individual, consolvo2021why}. 
Several studies reported that \users who are \cf{resource or time constrained} (emergency department workers~\cite{stobert2020understanding}, older adults~\cite{frik2019privacy,hornung2017navigating}, refugees~\cite{simko2018computer}, and others) did not understand \spasadj settings that were, in theory, available to them.

\subsection{Broken technology assumptions} 
\label{ss:broken-assumptions}

The atypical threat models \users face sometimes break assumptions that are built into secure system designs. Because of these assumptions, \spasadj best practices and technologies may be inaccessible, non-functional, or only minimally useful to \users, often with magnified consequences.

\paragraph{One person per device or account}
A common assumption of technology creators is that for every given device and account, there will be exactly one user who has access, despite prior work showing that convenience-based device sharing between trusted parties is common~\cite{matthews2016she}.\footnote{We note that since many of the papers in our dataset were published, there has been progress in addressing this issue, particularly in supporting child and family accounts. Nonetheless, we believe more can still be done to disrupt this assumption.} 
This core assumption fails for several \groups.
\Users who \cf{rely on a third party} may share account information with these trusted parties in order to accomplish important tasks~\cite{simko2018computer, kumar2019privacy} or receive needed monitoring~\cite{mentis2019upside}.
Different cultural privacy models have resulted in certain users sharing devices and accounts due to \cf{\socialnorms}~\cite{sambasivan2018privacy, ahmed2019everyone} or even the \cf{\political} requirements of where they live~\cite{alghamdi2015security}.
Additionally, \users who have a \cf{relationship with the attacker} were frequently forced to give these attackers access to their devices under duress, directly breaking this assumption~\cite{havron2019clinical, chatterjee2018spyware, freed2019my, freed2018stalker, chen2019computer, matthews2017stories}.

\paragraph{Everyone has sufficient technology access}
Some security techniques assume minimum levels of technology access that are out of reach for some at-risk users who are \cf{resource constrained}. For example, 2FA that depends on a mobile phone (e.g., via SMS or a code-generating app) may not be available to \users experiencing homelessness who are unable to consistently pay a phone bill or have an old device without space for new apps~\cite{sleeper2019tough}. 
Some low-income Black Americans reported needing to stay on family mobile plans---which could have enabled surveillance from untrusted relations---because they could not afford separate service~\cite{elliott2015:straighttalk}.
U.S. sanctions on Sudan entirely prevented Sudanese activists from enabling 2FA on a particular social media platform, since that platform's 2FA system was prohibited from recognizing Sudanese phone numbers~\cite{daffalla2021defensive}.

\paragraph{Physical and cognitive capacities are universal}
Even for general users, advice about authentication tends to assume 
impossible cognitive capabilities, such as memorizing a unique, strong 
password for every account~\cite{fbipasswords,nist800-63}.
This mistaken assumption is compounded in cases of \cf{underserved accessibility needs}. 
For example, several papers have identified password memorization as 
a critical challenge for older adults with cognitive impairments~\cite{mentis2019upside,frik2019privacy}.
Password memorization can be difficult for people with disabilities that relate to alphanumeric comprehension, like dyslexia or aphasia~\cite{marne2017learning}.

\paragraph{Concepts, values, and experiences are universal}
Other \spasadj paradigms rest on ideas, definitions, 
values, and morals that may not apply universally. 
Shortcomings in these assumptions can create \spasadj risks, especially for \users with different \cf{\socialnorms}. 
For example, translations from English to Khmer of concepts like \textit{privacy} within social media settings were hard to understand in the strongly community-oriented culture in Cambodia~\cite{jack2019privacy}. 
Similarly, some refugees who immigrated to the U.S. came from cultures where birthdays were not recorded; when these users were assigned a default value of January 1 in the U.S., 
authentication systems that relied on knowledge or entropic distribution of birthdays were less effective~\cite{simko2018computer}. Separately, Barwulor et al.\ \cite{barwulor2021disadvantaged} found that moral codes and laws enacted by the U.S. government and enforced by U.S. companies made it difficult for sex workers in other countries, where their 
work is legal, to access safe payment and advertisement platforms, 
placing their physical and digital safety at risk.

\paragraph{Limiting digital-safety options is good for everyone}
Digital-safety options that are too numerous can be hard to use~\cite{stanton2016security}.
But limiting the nuanced control enabled by \spasadj options to meet the usability needs of typical users may not always work well for \users. Risk factors that increased the chances of focused targeting---which included \cf{prominence}, \cf{relationship with the attacker}, \cf{reliance on a third party}, and \cf{access to other at-risk users}---can lead \users to have highly contextual \spasadj needs. For example, users protecting against a focused, intimate attacker, must account for nuances in the current relationship, the attacker's mood, whether or not they are physically copresent, and more~\cite{matthews2017stories,levy2020privacy}. Because of this nuance, \users may benefit from additional options or enhanced transparency that they can deploy in specific alignment with their goals. When these options are not easily available or understood, it may lead \users to fall back on \textit{distancing behaviors} in which they try to stay safe at the cost of fully engaging with technology~\cite{freed2017digital,lerner2020privacy, guberek2018keeping, mcgregor2015investigating,daffalla2021defensive,frik2019privacy}.

\section{Implications and Future Directions}
\label{s:future}

The at-risk framework can be used in multiple ways by researchers and technology creators, including guiding research and developing technologies to be inclusive of \users. 

\subsection{Research} 
The framework, as applied to our dataset in \autoref{tab:populations_to_factors}, can be used to help identify where knowledge of at-risk users is underdeveloped, sparse, or missing, giving researchers a way to prioritize their efforts.
The framework can also be used to guide development of study designs and research questions.

\paragraph{Identify the \textit{who} and the \textit{what}} 
The \spasadj community would benefit from research about all at-risk populations and factors---this includes expanding knowledge about those that have already been studied and 
creating new knowledge about those that have not. In this complex and 
nuanced space, even after several papers have reported on a population,
researchers may still have much to learn about the populations' \spasadj needs. 

Nonetheless, our meta-analysis makes clear that certain \groups and risk factors have received particularly limited attention, at least recently and within the \spasadj community. We observed a general tendency to study participants from Western cultures, especially the U.S.
\postsubmission{Studying people from other regions and cultures could shed light 
on new risks related to \cf{\socialnorms}, un- or understudied interactions among risk factors, or even new risk factors.}
We advocate for more research involving \postsubmission{geographically and culturally diverse at-risk participants, conducted by researchers who} represent the world. 

We also saw multiple studies exploring populations that experienced \cf{\marginalization}, focusing on a fairly narrow set of risk experiences (i.e., some populations' only black circle (\fullcirc) in \autoref{tab:populations_to_factors} was for the \cf{\marginalization} factor); future work could expand our understanding of \spasadj risks for these populations. Other groups often referenced as at-risk in popular media (e.g., real estate agents who have \cf{access to sensitive resources}~\cite{fbi-real-estate}; celebrities who face \spasadj risks due to \cf{prominence} and parasocial relationships~\cite{celebrity-parasocial}) have not been studied in the research venues we analyzed. New or additional foundational research may be needed to inform the \spasadj community about these populations' experiences and needs.

Our framework also highlights how risk factors often combine and interact, but the literature in our dataset has not deeply explored this topic, precluding a thorough synthesis. We advocate \textit{interactions} as a critical area of future research—one which our framework can support (e.g., enumerating the risk factors to consider for interactions). 

\postsubmission{Finally, in crisis situations, such as natural disasters or war, people may lose access to essential resources, their circumstances may change, or they may change their behaviors in exchange for critical services in ways that amplify their risk of attack or the severity of resulting harms. Recent papers on the impacts of the COVID-19 pandemic provide evidence that such changes can create new risks and amplify existing risks (\eg \cite{tseng2021covid,yamamoto2021teletherapy}). Future work on \spas during various kinds of crisis situations could yield a new risk factor, or expand the definitions of existing factors.}

\paragraph{Guide the \textit{how}} The framework can also be used to help 
shape study designs and reporting, particularly interview or survey questions. For a population about which little is known, asking about all 10 contextual risk factors can ensure fairly comprehensive coverage of digital-safety concerns. For populations where some research exists, researchers can use the framework to explore how previously un- or understudied risk factors may (or may not) apply, or add depth on the impact of a specific, previously identified risk factor of interest. \postsubmission{Researchers can also use the protective practices portion of the framework to explore more comprehensively how their participants currently protect themselves and why they choose their practices.} We hope that using the framework to guide research and reporting can enable better comparisons among studies, helping to uncover when disparate 
populations have overlapping (or distinctive) \spasadj practices and needs.
\postsubmission{Beyond this, it is important for researchers to carefully plan ethical methods when working with \users, guidance for which is an emerging area of research~\cite{liang2021embracing,costanza2018design}.}

\subsection{Technology development}

Technology creators can use the at-risk framework to better support a wide range of \users in their products.

\paragraph{Consider at-risk users at scale} \postsubmission{Our framework does not replace direct engagement with at-risk users, and we advocate for such engagement when appropriate. However, doing so selectively and ethically is important, and there are still open questions about ethical methods for at-risk user research (e.g., how do researchers not overtax already stressed and resource-constrained groups?). Meanwhile,} many papers in our dataset recommended considering the impact of a technology design on the at-risk population they studied, \postsubmission{which is incredibly important but} difficult to scale across populations without a guiding framework. \postsubmission{Our framework simplifies the challenging but important process of thinking through potential risks and needs of multiple at-risk populations together.} It does this by providing 10 contextual risk factors that organize patterns of risks and needs, \postsubmission{and a set of protective practices \users currently deploy to (sometimes ineffectively) cope. Using our framework can help researchers and technology creators prepare for user research, at interim points during multi-phased technology creation projects, or when user research across multiple at-risk populations is not an ethical option.} 

Each risk factor suggests specific, sometimes overlapping, technology needs. For \cf{\socialnorms}, \cf{relationship with the attacker}, and \cf{reliance on a third party}, users often could not keep their devices and accounts private. These users might benefit from the ability to keep select technology use and data secret on shared devices and accounts, or perhaps to enable digital traces that are ambiguous or imprecise to support plausible deniability~\cite{aoki2005making}.
Users with \cf{access to a sensitive resource} would benefit from robust protections for the sensitive resource (e.g., encryption, strong authentication). Users facing focused and/or sophisticated attackers (e.g., \cf{\political}, \cf{prominence}, \cf{relationship with the attacker}, \cf{\accessperson}) could benefit from easier-to-use versions of strong protections (such as hardware-based 2FA, strong passwords, and encryption) coupled with guided set-up flows and education. Users with \cf{prominence} or who experience \cf{\marginalization} would benefit from support for managing bulk or pervasive attacks from potentially anyone.

\postsubmission{Our systematization of protective practices (\autoref{s:protective}) and barriers (\autoref{s:risky}) together highlight how \users cope with their risks, sometimes in ways that are not completely effective or that introduce vulnerabilities. Notably, \users commonly employed non-technical practices---such as a host of social strategies and distancing behaviors---and it is important for technology creators to understand and not disrupt the important role these practices play. The protective practices we identified also show that \users are not commonly using some existing \spasadj solutions (at least as reported in the dataset), suggesting areas where technology creators could improve accessibility and/or usability. Further, we discuss broken assumptions that contribute to ineffective protections (\autoref{ss:broken-assumptions}), and encourage technology creators to consider these in new technologies.}

\paragraph{Balance tensions} Those creating technology for \users will have to contend with inherent tensions: ``perfect'' \spasadj protections usually do not exist. At-risk users already use a variety of practices, technical and otherwise, to address their pressing \spasadj concerns (\autoref{s:protective}). These practices all have some protective value, but may also come with significant downsides, like reduced social participation, lack of agency, and loss of transparency. For example, \emph{distancing} from technology (\autoref{ss:distancing}) was a common protective practice across our dataset, but it also reduced access to \emph{social support}, which was another essential protective practice broadly employed by at-risk users (\autoref{ss:social}). Technology creators should understand that any technical intervention is likely to introduce benefits and drawbacks, 
which should be carefully studied 
to help ensure such tensions are understood, manageable, and net beneficial. Careful evaluation of potential designs using our framework can help technology teams reason about how to balance various risks and benefits.

\paragraph{Balance usability and options}
\postsubmission{In \autoref{ss:broken-assumptions}, we discussed \users' need for more nuanced \spas options than typical users.} At the same time, adding options and transparency must be balanced with a very real need for usability. \Users experience heightened stress associated with higher potential for digital harm, as well as stress from their particular risk factors (\eg \cf{resource constraints}, \cf{\marginalization}, etc.). 

To balance the need for options that addresses the unique needs of \users with the high bar for usability, we encourage practitioners to make transparency and controls actionable and manageable. For example, transparency features that make users aware of a threat can be more actionable if they provide clear next steps to fix the issue and then guide the user through relevant protective measures. Similarly, layered or directed designs can help users find the options and controls that best meet their needs. Equally important are easy-to-understand defaults that minimize barriers to deploying protections and are carefully selected 
to support \users with many competing priorities (\autoref{ss:deliberate}). 

\section{Conclusion}
\label{s:conclusion}

Over the past several years, a growing body of research has focused on \spasadj risks for \users; however, guidance drawn from varied populations can be difficult for researchers and technology creators to apply in practice. To make this more tractable, we systematically analyzed \numpapers papers focused on varied populations and created an \emph{at-risk framework}:  10~\emph{contextual risk factors} that can augment or amplify common, high-priority \spasadj risks and their resulting harms, and the \emph{protective practices} \users employ to mitigate these risks. We used our framework to discuss \emph{barriers} \users face enacting digital protections. Going forward, our framework can be used to identify opportunities for future research and to provide a structure for researchers and technology creators to scalably and more comprehensively ensure that everyone---including \users---can engage safely online.

\section{Acknowledgements}
\label{s:acknowledgements}

\postsubmission{We thank our colleagues and experts who provided feedback on this work, including Allison McDonald, Andrew Botros, Beng Lim, Emerson Murphy-Hill, Florian Schaub, Franziska Roesner, Jill Palzkill Woelfer, Josh Lovejoy, Nafis Zebarjadi, Patrawat Samermit, Reena Jana, Shaun Kane, Stephan Somogyi, and Tu Tsao. 
This material is based upon work supported by DARPA under grant HR00112010011.  Any opinions, findings and conclusions or recommendations expressed in this material are those of the author(s) and do not necessarily reflect the views of the United States Government or DARPA.  Approved for public release; distribution is unlimited.}


{
\small
\bibliographystyle{plain}
\bibliography{libraryBib}
}

\vspace{60pt} 
\appendix
\hypertarget{}{}

\paragraph{Population categories and intersections}
\label{ss:pop_geo}
\postsubmission{Provisional categories~\cite{mccall2005intersectionality} of \groups represented by our dataset are listed in the first column of \autoref{tab:populations_to_factors}. While these population categories flatten some of the demographic richness reported in individual dataset papers, where possible we unpack intersectional issues using our thematic analysis and examples that include more detailed participant descriptions throughout the paper. For example, the ``activists'' category in \autoref{tab:populations_to_factors} describes activists across four continents, but examples note the specific geography when relevant. Further, these population categories are not exhaustive; instead they represent our dataset, demonstrating how risk factors can differ across categories and intersect for a single category.}

\paragraph{Biases in geographic representation}
\postsubmission{\Groups are found around the world, with social and structural circumstances that vary globally. The papers identified in our review skewed heavily toward Western, and specifically U.S., populations, but do include some studies with participants from other regions. About half of the papers in our dataset (48 out of \numpapers papers) included participants from the U.S. only, and 68\% of papers included participants from Western countries only. Populations about which the literature reported perspectives from around the world (including most or all continents), included activists, journalists, NGO staff, crowdworkers, LGBTQ+ people, and survivors of trafficking. Other non-Western perspectives were reported for women in South Asia, older adults in Cambodia, people in developing regions (in Asian and African countries), refugees (resettled from Africa and Asia to the US), and people with visual impairments in India. The remaining populations listed in \autoref{tab:populations_to_factors} represent Western perspectives.}

\paragraph{Researcher reflexivity}
\postsubmission{Though we do not know how the authors of dataset papers would identify their nationalities or describe their lived experiences, the majority appeared to be completed by teams at institutions in Western contexts. Similarly, we work in U.S.-based organizations and have backgrounds in HCI, computer science, security, and/or privacy. These contexts influenced the papers we sampled and the analysis lenses used---both by our team and the researchers who produced dataset papers. While this SoK establishes a framework, we advocate for additional perspectives to enrich our findings.}

\end{document}